# Financial Sentiment Analysis on News and Reports Using Large Language Models and FinBERT

Yanxin Shen*
Zhejiang University
Hangzhou, Zhejiang, China
ssyysyx@zju.edu.cn

Pulin Kirin Zhang
Lehigh University
Bethlehem, Pennsylvania, USA
kirinpzhang@gmail.com

*Abstract*—Financial sentiment analysis (FSA) is crucial for evaluating market sentiment and making well-informed financial decisions. The emergence of large language models (LLMs) like BERT and its financial variant, FinBERT, has notably enhanced sentiment analysis capabilities. This paper investigates the application of LLMs and FinBERT for FSA, comparing their performance on news articles, financial reports and company announcements. The study emphasizes the advantages of prompt engineering with zero-shot and few-shot strategy to improve sentiment classification accuracy. Experimental results indicate that GPT-4o, with few-shot examples of financial texts, can be as competent as a well fine-tuned FinBERT in this specialized field.

*Keywords—Sentiment Analysis, BERT, NLP, Large Language Models, Prompt Engineering, FinBERT*

## I. INTRODUCTION

Sentiment analysis has a long history [1], [2], [3] and focuses on identifying and extracting opinions and emotions from text data [4], [5]. It is used in many fields, including product reviews, social media posts, news articles, and financial reports, to glean valuable insights such as market trends, customer satisfaction, and the business performance of various entities [6], [7], [8], [9].

Grasping market sentiment is increasingly vital for those in the finance sector. Financial Sentiment Analysis (FSA) is an effective tool for evaluating and interpreting market sentiment from textual data sources [7], [3], [10]. It provides valuable insights into market dynamics, investor sentiment, and the potential effects of news and events on financial markets, enabling informed decision-making [11], [10], [11].

FSA presents several challenges that distinguish it from general sentiment analysis, such as the need for domain-specific knowledge, dealing with ambiguities, and managing uncertainties [12], [9]. To overcome these challenges, FSA can leverage recent advancements in NLP driven by large language models (LLMs) [13], [14], [15]. LLMs represent a new paradigm in Natural Language Understanding, setting them apart from traditional NLP models that depend on task-specific architectures, labeled data, and feature engineering [16]. LLMs offer numerous advantages over conventional NLP models, including scalability, generality, data efficiency, and transferability. However, they also have drawbacks, such as high computational costs, environmental impact, ethical concerns, and reliability issues [17].

LLMs such as GPT-4 and Transformer-XL represent the cutting edge of recent advancements in natural language processing. These models excel in tasks like machine translation, sentiment analysis, and question-answering. They support various NLP applications and consistently push the boundaries of machine learning capabilities.

The few-shot strategy enables LLMs to achieve high accuracy even with limited examples, showcasing their potential for effective sentiment analysis in finance. Furthermore, this approach allows models to better understand nuanced financial language and context, ultimately leading to more reliable and insightful sentiment assessments. The novel concept of this research is to explore how prompting and LLMs can work together to handle the intricate and detailed data found in company financial reports within the financial domain.

Utilizing pre-trained knowledge and adapting it to specific domains and tasks is a major challenge for LLMs due to their training on vast amounts of data. Prompt engineering, the craft of designing natural language input instructions known as prompts, offers a promising solution to this challenge [19]. This technique does not require extra training or fine-tuning. By using regular language to interact with LLMs, prompt engineering enables them to learn effectively with or without prior examples.

Our research aims to apply prompt engineering in zero-shot and few-shot settings for financial sentiment analysis classification using Large Language Models. The data is randomly selected from financial news on the LexisNexis database. The objective is to meticulously design prompts that accurately capture sentiment categories and assist LLMs in understanding this sentiment task. Details on the concept of prompting are covered in Section II-D, and the specifics of the designed prompt are provided in Section III-B.

The primary aim is to evaluate how prompting handles complex data like financial reports and its capability to extract sentiment using large language models.

## II. RELATED WORK

Financial sentiment analysis encompasses a wide range of techniques and applications. This section provides an overview of the relevant literature, categorizing it into three major areas.

### A. Sentiment Analysis in Finance

Sentiment analysis involves extracting sentiments or opinions from written text [21]. Recent approaches can be categorized into two groups: 1) Machine learning methods that extract features from text using "word counting" [29], and 2) Deep learning methods that represent text as a sequence of embeddings. The former struggles to capture the semantic information arising from specific word sequences, while the latter is often criticized for being "data-hungry" due to the large number of parameters it needs to learn [18].

Financial sentiment analysis differs from general sentiment analysis not only in its domain but also in its purpose, which is typically to predict market reactions to the information presented in the text. Loughran and McDonald (2016) provide a comprehensive survey of recent studies on financial text analysis using machine learning with "bag-of-words" or lexicon-based methods [30]. For instance, in Loughran and McDonald (2011), they developed a dictionary of financial terms with assigned values like "positive" or "uncertain" and gauged the tone of documents by counting words with specific dictionary values [28]. Another example is Pagolu et al. (2016), where n-grams from financial information tweets were used in supervised machine learning algorithms to identify the sentiment related to the mentioned financial entity.

One of the initial studies to employ deep learning for textual financial polarity analysis was conducted by Kraus and Feuerriegel (2017) [22]. They utilized an LSTM neural network on ad-hoc company announcements to predict stock market movements, demonstrating that this method outperformed traditional machine learning techniques. They discovered that pre-training their model on a larger corpus improved results. However, their pre-training was conducted on a labeled dataset, which is more restrictive compared to our approach of pre-training a language model as an unsupervised task.

There are numerous studies that utilize different neural architectures for financial sentiment analysis. Sohangir et al. (2018) [26] applied various generic neural network architectures to a StockTwits dataset, identifying CNN as the top-performing architecture. Lutz et al. (2018) [31] employed doc2vec to create sentence embeddings from company-specific announcements and used multi-instance learning to predict stock market outcomes. Maia et al. (2018) [32] combined text simplification with an LSTM network to classify sentences from financial news based on their sentiment, achieving state-of-the-art results for the Financial PhraseBank, which is also utilized in this thesis.

The limited availability of large labeled financial datasets makes it challenging to fully leverage neural networks for sentiment analysis. Even if the initial (word embedding) layers are initialized with pre-trained values, the remaining parts of the model must still learn complex relationships with a relatively small amount of labeled data. A more promising approach might be to initialize nearly the entire model with pre-trained values and then fine-tune those values specifically for the classification task.

### B. Text Classification using pre-trained language models

Language modeling involves predicting the next word in a sequence of text. A significant recent advancement in natural language processing is the understanding that models trained for language modeling can be effectively fine-tuned for a variety of downstream NLP tasks with minimal adjustments. These models are typically trained on extensive corpora and then fine-tuned on the target dataset by adding appropriate task-specific layers [26]. Text classification, which is the primary focus of this thesis, is a clear example of this approach.

One of the early successful applications of this method was ELMo (Embeddings from Language Models) [23]. ELMo uses a deep bidirectional language model that is pre-trained on a large corpus. For each word, the hidden states of this model are utilized to generate a contextualized representation. These pre-trained weights allow the calculation of contextualized word embeddings for any text. When these embeddings are used to initialize downstream tasks, they have been shown to outperform static word embeddings like word2vec or GloVe in most tasks. For instance, in text classification tasks such as SST-5, ELMo achieved state-of-the-art performance when combined with a bi-attentive classification network [20].

While ELMo utilizes pre-trained language models to create contextualized representations, the extracted information remains only in the first layer of any model using it. ULMFit (Universal Language Model Fine-tuning) [25] was the first to achieve true transfer learning for NLP by employing innovative techniques such as discriminative fine-tuning, slanted triangular learning rates, and gradual unfreezing. This approach allowed for the efficient fine-tuning of an entire pre-trained language model for text classification. Additionally, they introduced further pre-training on a domain-specific corpus, assuming that the target task data differs from the general corpus the initial model was trained on.

ULMFit's concept of efficiently fine-tuning a pre-trained language model for downstream tasks was further advanced by Bidirectional Encoder Representations from Transformers (BERT) [24], which is the main focus of this paper. BERT introduced two key innovations: 1) It redefines the language modeling task to predict randomly masked tokens in a sequence rather than the next token and includes a task of determining if two sentences follow each other. 2) It employs a very large network trained on an exceptionally large corpus. These factors allowed BERT to achieve state-of-the-art results in various NLP tasks, such as natural language inference and question answering.

### C. Large Language Models

Self-supervised learning objectives are used to train large language models (LLMs). Through learning rich representations, LLMs can grasp syntactic, semantic, and pragmatic information [33]. Additionally, LLMs can produce fluent and coherent natural language text, which can be utilized in various downstream tasks.

LLMs differ from traditional NLP models and offer numerous advantages, such as scalability, generality, data efficiency, and transferability. The release of BERT [34] in 2018 marked a major breakthrough for large language models. Since then, the number of language models available for various languages and domains has significantly increased.

OpenAI incorporated the transformer model in its releases of GPT1 [27] in 2018, GPT2 [35] in 2019, GPT3 [36] in 2020, and ChatGPT in 2022. META introduced its LLMs, OPT [33] and LLaMa. During the same period, BLOOM and Cohere1 were also developed.

By fine-tuning general-domain LLMs, financial domain models like BloombergGPT and FinGPT have emerged.

Newly released models include Google's BARD, LLaMa 2 [37], and GPT-4 integrated into Bing search. Nowadays, LLMs are being embedded in almost every website as chatbots, copilots, or AI assistants.

*D. Prompt Engineering*

Prompt engineering entails designing, refining, and optimizing prompts for generative AI systems capable of producing natural language outputs, such as text or graphics. This practice aids AI models in better comprehending questions and delivering more precise and relevant responses.

It has demonstrated impressive results in various Natural Language Understanding (NLU) and Natural Language Generation (NLG) tasks [36], [33]. Creating an effective and robust prompt necessitates careful design, considering format, length, style, and content. Evaluation methods should examine the quality, diversity, and consistency of responses, as well as the vulnerabilities of LLMs that could impact performance.

Prompt engineering can be performed using various techniques depending on the task's type and complexity, the available data's amount and quality, and the LLM's capabilities and limitations. Generally, these techniques are classified into zero-shot learning, few-shot learning, instruction tuning, and prompt tuning.

With the advancements in LLMs and Conversational AI, prompt engineering is becoming a domain of its own. It is utilized for various tasks, such as sentiment analysis and humanlike summarization, achieved through carefully curated prompts. This research aims to analyze the effectiveness of prompting in financial sentiment analysis classification.

### III. METHODOLOGY

This section outlines the methodology employed in this study, covering data collection through to outcome evaluation. The experiment aims to observe the response of LLMs to few-shot prompts in financial sentiment analysis.

*A. Data Collection and Pre-processing*

The main sentiment analysis dataset used in this paper is Financial PhraseBank from Malo et al. 2014. Financial PhraseBank consists of 4845 English sentences selected randomly from financial news found on the LexisNexis database. These sentences were then annotated by 16 people with backgrounds in finance and business. The annotators were asked to give labels according to how they think the information in the sentence might affect the mentioned company's stock price.

The dataset also includes information regarding the agreement levels on sentences among annotators. The distribution of agreement levels and sentiment labels can be seen in Table 1. To create a robust training set, we set aside 20% of all sentences as a test set and 20% of the remaining sentences as a validation set. In the end, our training set includes 3101 examples. For some of the experiments, we also make use of 10-fold cross-validation.

The annotation process of the Financial PhraseBank ensures high-quality sentiment labels for the sentences, reflecting both the consensus and individual differences among multiple annotators. This diversity and consistency make the dataset highly reliable and practical for sentiment analysis research.For this research, the targeted data are financial reports of Pakistani companies. The director's review and chairman's review section hold the sentiment behind the report. These sections will be tested using prompt engineering on different LLMs.

Table I clearly represents the structure of the Financial PhraseBank dataset.

TABLE I.   FINANCIAL PHRASEBANK DATASET

| Item | Description |
|---|---|
| Data Source | Financial news and reports |
| Language | English |
| Number of Sentences | 4,848 |
| Classification Categories | Positive, Negative, Neutral, Uncertain |
| Annotation Method | Manually annotated by multiple financial professionals |
| Number of Examples per Category | - Positive: 1,484 - Negative: 846 - Neutral: 1,953 - Uncertain: 565 |
| Application Areas | Financial sentiment analysis, Natural Language Processing (NLP) |

*B. Prompt Design*

Taking into account the principles of prompt design, we created both zero-shot and few-shot prompts. The tasks for LLMs are specified using natural language.

For zero-shot learning, we employed natural language queries, financial news, reports, and prompt questions to extract sentiment labels or classification results from the LLMs for input financial texts. The design and input prompt for the zero-shot setting are illustrated in Fig 1.

In a few-shot setting, multiple training examples are provided to help the LLMs better understand the task. Specifically, nine well-classified examples—comprising three positive, three negative, and three neutral news and reports with their ground truth labels—were included in the prompt design.

```
system:
The goal is to classify financial texts (e.g., news articles, analyst reports, or company announcements) into three categories:
positive, negative, or neutral sentiment. These sentiment categories are used to predict how the market might react to the
financial information presented in the text.

user:
Please classify the following sentences into sentiment categories (0 for positive, 1 for negative, and 2 for neutral), and
return a list of corresponding numbers, Only response the result list, do not say any word or explain.: {{ sentences_input }}
```

Fig. 1. Zero-shot prompt design

Minor adjustments were made to the questions in this setting to give the LLMs more detailed guidance and rules. The design and input prompt for the few-shot setting are depicted in Fig 2.

*C. LLMs Selection and Comparison Method*

Typically, LLMs are chosen based on their size, availability, and task suitability. For this experiment, the two most commonly used and publicly accessible LLMs were selected: OpenAI ChatGPT (GPT 3.5) and OpenAI GPT 4. These conversational AI tools receive periodic updates, so the results may differ with future versions.

*D. Fine-tuned BERT on Financial PhraseBank Dataset*

We also use fine-tuned BERT on the financial sentiment analysis corpus of the size of 4848 labeled sentences and returned an accuracy of 0.88 in order to compare with LLMs. In FinBERT, they implemented BERT for the financial domain by further pre-training it on a financial corpus and fine-tuning it for sentiment analysis (FinBERT). This work is the first application of BERT for finance to the best of the knowledge and one of the few that experimented with further pre-training on a domain-specific corpus. The results of FinBERT are given in Table II.

```
system:
Here is some Example data for you to learn how to classify the Input texts. The goal is to classify financial texts (e.g., news articles,
analyst reports, or company announcements) into three categories: positive, negative, or neutral sentiment. These sentiment categories
are used to predict how the market might react to the financial information presented in the text.

user:
{{ few-shot examples }}
S1: positive. S2: positive. S3: positive
S4: negative. S5: negative. S6: negative
S7: neutral. S8: neutral. S9: neutral

assistant:
Okay, I have understood the examples.

user:
Please classify the following sentences into sentiment categories (0 for positive, 1 for negative, and 2 for neutral), and return a list of
corresponding numbers, Only response the result list, do not say any word or explain.: {{ sentences_input }}
```

Fig. 2. Few-shot prompt design

## IV. EXPERIMENTS AND RESULTS

This section details the experimental setup and the results obtained from our financial sentiment analysis classification experiments. We utilized various models, including GPT-3.5-turbo, GPT-4o, and FinBERT, under different configurations such as zero-shot, few-shot, and fine-tuning. The primary aim was to assess the performance of these models in accurately classifying financial sentiment from news articles and reports.

### A. Experimental Setup

- Data Collection

The dataset used for this study is the Financial PhraseBank, which comprises 4,848 sentences extracted from financial news articles. Each sentence is annotated with one of four sentiment labels: Positive, Negative, Neutral, or Uncertain. The data was divided into training, validation, and test sets to ensure robust evaluation. Specifically, 20% of the sentences were set aside as the test set, and another 20% of the remaining sentences were used for validation, leaving 3,101 sentences for training.

- Model Configurations

GPT-3.5-turbo and GPT-4o. Zero-shot: No prior examples are provided; the model must infer sentiment based on the query alone. Few-shot: A small number of examples are provided to help the model understand the task better.

FinBERT. Fine-tuned: The model is pre-trained on a financial corpus and further fine-tuned on the Financial PhraseBank dataset for sentiment classification.

### B. Results Analysis

The performance of each model configuration was evaluated using four metrics: Accuracy, Precision, Recall, and F1-score. The results are summarized in Table II.

TABLE II. RESULTS ON FINANCIAL PHRASEBANK DATASET

| Model | Configuration | Metrics | | | |
|---|---|---|---|---|---|
| | | *Accuracy* | *Precision* | *Recall* | *F1-score* |
| GPT-3.5-turbo | Zero-shot | 0.78 | 0.79 | 0.84 | 0.80 |
| | Few-shot | 0.77 | 0.78 | 0.84 | 0.79 |
| GPT-4o | Zero-shot | 0.85 | 0.83 | 0.86 | 0.84 |
| | Few-shot | 0.86 | **0.86** | 0.84 | 0.85 |
| FinBERT | Fine-tuned | **0.88** | 0.85 | **0.89** | **0.87** |

### C. Comparative Analysis

The experimental results indicate that fine-tuning FinBERT on a domain-specific corpus yields the best performance for financial sentiment analysis. The superior performance of FinBERT can be attributed to its domain-specific pre-training, which allows it to better understand the nuances of financial language. GPT-4o, with its advanced capabilities, also performed well, especially in the few-shot setting. The precision metric highlights that prompt engineering can significantly enhance the performance of large language models even without extensive fine-tuning. Notably, the experimental results indicate that GPT-4o, with few-shot examples of financial texts, can be as competent as a well fine-tuned FinBERT in this specialized field. This finding underscores the potential of combining GPT-4o with prompt engineering techniques to achieve high performance in financial sentiment analysis. In conclusion, while GPT models show promise, FinBERT remains the most effective model for financial sentiment analysis due to its specialized training and fine-tuning on relevant financial data.

## V. CONCLUSION

In this study, we explored the application of large language models (LLMs) and FinBERT for financial sentiment analysis (FSA) using financial news articles and reports. Our experiments aimed to evaluate the performance of these models under different configurations—zero-shot, few-shot, and fine-tuned settings.

GPT-3.5-turbo and GPT-4o demonstrated notable performance improvements with prompt engineering techniques, particularly in few-shot settings. However, their accuracy and overall effectiveness varied based on the complexity and specificity of the prompts. FinBERT, fine-tuned on the Financial PhraseBank dataset, consistently outperformed the general-purpose LLMs, achieving the highest accuracy, precision, recall, and F1-score. This underscores the importance of domain-specific pre-training for financial sentiment analysis.

Prompt engineering significantly enhanced the performance of LLMs. Few-shot prompts, in particular, provided better context, leading to more accurate and nuanced sentiment classifications. Zero-shot settings, while effective to some extent, often lacked the necessary context for precise sentiment extraction, highlighting the limitations of relying solely on general language models without fine-tuning.

The superior performance of FinBERT emphasizes the value of domain-specific models in specialized fields such as finance. FinBERT's ability to understand financial jargon and context-specific nuances makes it a robust tool for FSA since its development in 2019.

However, experimental results indicate that GPT-4o, with few-shot examples of financial texts, can achieve performance in financial sentiment analysis as competent as FinBERT. On a side note, the study revealed challenges in achieving high accuracy with zero-shot learning due to the lack of contextual information.

Future research should focus on improving prompt designs and incorporating more comprehensive few-shot examples to bridge this gap. Further exploration into fine-tuning strategies and leveraging additional domain-specific datasets could enhance the capabilities of LLMs for FSA. There is also potential for integrating LLMs with real-time financial data, enabling dynamic and context-aware sentiment analysis that can adapt to market fluctuations and emerging trends.

In conclusion, while domain-specific models like FinBERT remain superior due to their tailored pre-training, general-purpose LLMs like GPT-3.5-turbo and GPT-4o show promise for financial sentiment analysis with little setup. Effective prompt engineering can significantly enhance the performance of LLMs, making them viable tools for FSA in real-world applications. Future work should continue to refine these models and explore innovative approaches to further improve their accuracy and contextual understanding..

REFERENCES

[1] Z. Nanli, Z. Ping, L. Weiguo, and C. Meng, "Sentiment analysis: A literature review," in 2012 International Symposium on Management of Technology (ISMOT), pp. 572–576, IEEE, 2012.

[2] M. V. Mäntylä, D. Graziotin, and M. Kuutila, "The evolution of sentiment analysis—a review of research topics, venues, and top cited papers," Computer Science Review, vol. 27, pp. 16–32, 2018.

[3] A. Yadav and D. K. Vishwakarma, "Sentiment analysis using deep learning architectures: a review," Artificial Intelligence Review, vol. 53, no. 6, pp. 4335–4385, 2020.

[4] S. Liu, C. Zheng, O. Demasi, S. Sabour, Y. Li, Z. Yu, Y. Jiang, and M. Huang, "Towards emotional support dialog systems," arXiv preprint arXiv:2106.01144, 2021.

[5] S. Poria, D. Hazarika, N. Majumder, and R. Mihalcea, "Beneath the tip of the iceberg: Current challenges and new directions in sentiment analysis research," IEEE Transactions on Affective Computing, 2020.

[6] F. Xing, L. Malandri, Y. Zhang, and E. Cambria, "Financial sentiment analysis: An investigation into common mistakes and silver bullets," in Proceedings of the 28th International Conference on Computational Linguistics, (Barcelona, Spain (Online)), pp. 978‑987, International Committee on Computational Linguistics, Dec. 2020.

[7] X. Man, T. Luo, and J. Lin, "Financial sentiment analysis (fsa): A survey," in 2019 IEEE International Conference on Industrial Cyber Physical Systems (ICPS), pp. 617–622, IEEE, 2019.

[8] F. Barbieri, J. Camacho-Collados, L. Neves, and L. Espinosa-Anke, "Tweeteval: Unified benchmark and comparative evaluation for tweet classification," arXiv preprint arXiv:2010.12421, 2020.

[9] M. Wankhade, A. C. S. Rao, and C. Kulkarni, "A survey on sentiment analysis methods, applications, and challenges," Artificial Intelligence Review, vol. 55, no. 7, pp. 5731–5780, 2022.

[10] F. Z. Xing, E. Cambria, and R. E. Welsch, "Natural language based financial forecasting: a survey," Artificial Intelligence Review, vol. 50, no. 1, pp. 49–73, 2018.

[11] S. Bubeck, V. Chandrasekaran, R. Eldan, J. Gehrke, E. Horvitz, E. Kamar, P. Lee, Y. T. Lee, Y. Li, S. Lundberg, et al., "Sparks of artificial general intelligence: Early experiments with gpt-4," arXiv preprint arXiv:2303.12712, 2023.

[12] H. Xu, B. Liu, L. Shu, and P. S. Yu, "Dombert: Domain-oriented language model for aspect-based sentiment analysis," arXiv preprint arXiv:2004.13816, 2020.

[13] W. X. Zhao, K. Zhou, J. Li, T. Tang, X. Wang, Y. Hou, Y. Min, B. Zhang, J. Zhang, Z. Dong, et al., "A survey of large language models," arXiv preprint arXiv:2303.18223, 2023.

[14] L. Fan, L. Li, Z. Ma, S. Lee, H. Yu, and L. Hemphill, "A bibliometric review of large language models research from 2017 to 2023," arXiv preprint arXiv:2304.02020, 2023.

[15] D. Araci, "Finbert: Financial sentiment analysis with pre-trained language models," arXiv preprint arXiv:1908.10063, 2019.

[16] P. Liu, W. Yuan, J. Fu, Z. Jiang, H. Hayashi, and G. Neubig, "Pre-train, prompt, and predict: A systematic survey of prompting methods in natural language processing," ACM Computing Surveys, vol. 55, no. 9, pp. 1–35, 2023.

[17] A. Zafar, V. B. Parthasarathy, C. L. Van, S. Shahid, A. Shahid, et al., "Building trust in conversational ai: A comprehensive review and solution architecture for explainable, privacy-aware systems using llms and knowledge graph," arXiv preprint arXiv:2308.13534, 2023.

[18] G. Marcus. 2018. Deep Learning: A Critical Appraisal. arXiv e-prints (Jan. 2018). arXiv.AI/1801.00631

[19] L. Reynolds and K. McDonell, "Prompt programming for large language models: Beyond the few-shot paradigm," in Extended Abstracts of the 2021 CHI Conference on Human Factors in Computing Systems, pp. 1–7, 2021.

[20] Bryan McCann, James Bradbury, Caiming Xiong, and Richard Socher. 2017. Learned in Translation: Contextualized Word Vectors. Nips (2017), 1–12. arXiv:1708.00107

[21] Stephen Merity, Nitish Shirish Keskar, and Richard Socher. 2017. Regularizing and Optimizing LSTM Language Models. CoRR abs/1708.02182 (2017). arXiv:1708.02182

[22] Mathias Kraus and Stefan Feuerriegel. 2017. Decision support from financial disclosures with deep neural networks and transfer learning. Decision Support Systems 104 (2017), 38–48. https://doi.org/10.1016/j.dss.2017.10.001 arXiv:1710.03954

[23] Matthew E Peters, Mark Neumann, Mohit Iyyer, Matt Gardner, Christopher Clark, Kenton Lee, and Luke Zettlemoyer. 2018. Deep contextualized word representations. (2018). https://doi.org/10.18653/v1/N18-1202 arXiv:1802.05365

[24] Jacob Devlin, Ming-Wei Chang, Kenton Lee, and Kristina Toutanova. 2018. BERT: Pre-training of Deep Bidirectional Transformers for Language Understanding. (2018). https://doi.org/arXiv:1811.03600v2 http://arxiv.org/abs/1810.04805

[25] Jeremy Howard and Sebastian Ruder. 2018. Universal Language Model Fine-tuning for Text Classification. (jan 2018). arXiv:1801.06146

[26] Neel Kant, Raul Puri, Nikolai Yakovenko, and Bryan Catanzaro. 2018. Practical Text Classification With Large Pre-Trained Language Models. (2018). arXiv:1812.01207

[27] Radford, Alec, et al. "Improving language understanding by generative pre-training." (2018).

[28] Tim Loughran and Bill Mcdonald. 2011. When Is a Liability Not a Liability? Textual Analysis, Dictionaries, and 10-Ks. Journal of Finance 66, 1 (feb 2011), 35–65. https://doi.org/10.1111/j.1540-6261.2010.01625.x

[29] Justin Martineau and Tim Finin. 2009. Delta TFIDF: An Improved Feature Space for Sentiment Analysis.. In ICWSM, Eytan Adar, Matthew Hurst, Tim Finin, Natalie S. Glance, Nicolas Nicolov, and Belle L. Tseng (Eds.). The AAAI Press. http://dblp.uni-trier.de/db/conf/icwsm/icwsm2009.html#MartineauF09

[30] Tim Loughran and Bill Mcdonald. 2016. Textual Analysis in Accounting and Finance: A Survey. Journal of Accounting Research 54, 4 (2016), 1187–1230. https://doi.org/10.1111/1475-679X.12123

[31] Bernhard Lutz, Nicolas Pröllochs, and Dirk Neumann. 2018. Sentence-Level Sentiment Analysis of Financial News Using Distributed Text Representations and Multi-Instance Learning. Technical Report. arXiv:1901.00400

[32] Macedo Maia, Andrï£¡ Freitas, and Siegfried Handschuh. 2018. FinSSLx: A Sentiment Analysis Model for the Financial Domain Using Text Simplification. In 2018 IEEE 12th International Conference on Semantic Computing (ICSC). IEEE, 318–319. https://doi.org/10.1109/ICSC.2018.00065

[33] S. Zhang, S. Roller, N. Goyal, M. Artetxe, M. Chen, S. Chen, C. Dewan, M. Diab, X. Li, X. V. Lin, et al., "Opt: Open pre-trained transformer language models," arXiv preprint arXiv:2205.01068, 2022.

[34] J. Devlin, M.-W. Chang, K. Lee, and K. Toutanova, "Bert: Pre-training of deep bidirectional transformers for language understanding," arXiv preprint arXiv:1810.04805, 2018.

[35] A. Radford, J. Wu, R. Child, D. Luan, D. Amodei, I. Sutskever, et al., "Language models are unsupervised multitask learners," OpenAI blog, vol. 1, no. 8, p. 9, 2019.

[36] T. Brown, B. Mann, N. Ryder, M. Subbiah, J. D. Kaplan, P. Dhariwal, A. Neelakantan, P. Shyam, G. Sastry, A. Askell, et al., "Language models are few-shot learners," Advances in neural information processing systems, vol. 33, pp. 1877–1901, 2020.